\documentstyle[preprint,aps]{revtex}
\draft

\newcommand{\ba}{\begin{eqnarray}}
\newcommand{\ea}{\end{eqnarray}}

\begin{document}
\bibliographystyle{plain}
\title
{Partial Dynamical Symmetry and Mixed Dynamics}
\author
{A. Leviatan$^{a}$ and N. D. Whelan$^{b,c}$}
\address{$^{a}$Racah Institute of Physics,
The Hebrew University, Jerusalem 91904, Israel
\newline
$^{b}$Centre for Chaos and Turbulence Studies, Niels Bohr Institute,
Copenhagen, Denmark
\newline
$^{c}$Division de Physique Th\'eorique\cite{cnrsbullshit},
IPN, 91406 Orsay Cedex, France}


\maketitle

\begin{abstract}
Partial dynamical symmetry describes a situation in which some
eigenstates have a symmetry which the quantum Hamiltonian does not share.
This property is shown to have a classical analogue in which some tori in phase
space are associated with a symmetry which the classical Hamiltonian
does not share. A local analysis in the vicinity of these special tori reveals
a neighbourhood of phase space foliated by tori. This clarifies the
suppression of classical chaos associated with partial dynamical symmetry.
The results are used to divide the states of a mixed system into ``chaotic''
and ``regular'' classes.
\end{abstract}

\pacs{PACS numbers: 05.45.+b, 03.65.Fd, 03.65.Sq, 11.30.Na}

Symmetry plays a central role in affecting the character of classical and
quantum dynamics. Constants of motion associated with a symmetry
govern the integrability of the system. Often
the symmetry in question is not obeyed uniformly,
i.e. only a subset of quantum states fulfill the symmetry
while other states do not. In such circumstances, referred to as partial
symmetries, the symmetry of these ``special'' (at times solvable)
eigenstates does not arise from invariance properties of the Hamiltonian.
Selected examples in this category are
adiabatic regular states in the stadium billiard \cite{quasiad};
regular quasi-Landau resonances of a hydrogen
atom in strong magnetic fields
\cite{quasila}; discrete nuclear states
embedded in a continuum of decay channels \cite{rotter};
partial $SU(3)$ symmetry found in deformed nuclei \cite{al96}.
Hamiltonians with partial symmetries are not completely integrable hence
can exhibit stochastic behavior. As such they are relevant
to the study of mixed systems with coexisting regularity and chaos
\cite{mixed}, which are the most generic.

The effect of discrete symmetries on a mixed-phase-space system was examined
in \cite{btu90}, while the consequences of breaking a discrete symmetry
on the spectral statistics of a chaotic Hamiltonian were discussed in
\cite{lkc94}. In the present paper we focus on continuous symmetries
(associated with Lie groups) which can be conveniently studied in the
framework of algebraic models. Such symmetry-based models
are amenable to both quantum and classical treatments and
have been used extensively in nuclear and molecular physics \cite{ibm}.
Their integrable limits are associated with dynamical
symmetries in which the Hamiltonian is written in terms of Casimir
operators of a chain of nested algebras. The eigenstates and eigenvalues
are labeled by the irreducible representations (irreps) of the 
algebras in the chain, and are known analytically. Both the quantum
and classical Hamiltonians are completely solvable \cite{ibmchaos}.
Non-integrability is obtained by breaking the dynamical symmetry and
may lead to chaotic dynamics \cite{ibmchaos,mere}.
To address situations of mixed dynamics,
an algorithm was developed \cite{al} for constructing algebraic Hamiltonians
with partial dynamical symmetries. Such Hamiltonians are not invariant
under a symmetry group and yet possess a subset of ``special'' solvable states
which do respect the symmetry.
In the context of a nuclear physics model
involving five quadrupole degrees of freedom, it was shown that partial
dynamical symmetry (pds) induced a strong suppression of classical chaos
\cite{wal}. This was true even though the fraction of special
states vanished as $\hbar^2$, so one might have expected no
classical effect. In order to better understand this effect, we consider
a simpler model and use its pds
to infer relationships between the classical
and quantum dynamics of a Hamiltonian in a mixed KAM r\'egime.

As a simple test-case we consider a model based on a $U(3)$ algebra.
This algebra (with fermionic operators) was considered previously in the
context of chaos, but not regarding pds
\cite{mere}.
Here we employ a realization in terms of three types of bosons
$a^{\dagger}$, $b^{\dagger}$, $c^{\dagger}$ satisfying the usual
commutation relations.
The nine number-conserving bilinear products of creation and destruction
operators comprise the U(3) algebra. The conservation
of the total boson-number $\hat{N}=\hat{n}_a+\hat{n}_{b}+\hat{n}_c$
($\hat{n}_a =a^{\dagger}a$ with eigenvalue $n_a$ etc.) 
ensures that the model describes a system
with only two independent degrees of freedom. All states of the model are
assigned to the totally symmetric representation [N] of $U(3)$.
One of the dynamical symmetries of the model is associated with the
following chain of algebras
\begin{equation} \label{chain}
U(3) \supset U(2) \supset U(1)
\end{equation}
Here $U(2)\equiv SU(2)\times U_{ab}(1)$ with a linear Casimir
$\hat{n}_{ab}=\hat{n}_a+\hat{n}_b$ (which is also the generator of
$U_{ab}(1)$ ). The generators of $SU(2)$ are $\hat{J}_+ = b^{\dagger}a$,
$\hat{J}_-= a^{\dagger}b$, $\hat{J}_z=(\hat{n}_b-\hat{n}_a)/2$ and its
Casimir $\vec{J}^2=\hat{n}_{ab}(\hat{n}_{ab}+2)/4$. The subalgebra $U(1)$
in Eq. (\ref{chain}) is composed of the operator $\hat{J}_z$.
A choice of Hamiltonian with a $U(2)$ dynamical symmetry is
\begin{eqnarray}
H_0 & = &
\omega_a a^{\dagger}a + \omega_b b^{\dagger}b
\;\; = \;\;
\hat{n}_{ab} - 2A\hat{J}_z
\label{h0}
\end{eqnarray}
where $\omega_{a,b} = 1 \pm A$, and $A$ is introduced to break degeneracies.
Diagonalization of this Hamiltonian is trivial and leads to eigenenergies
$E_{n_a,n_b}= \omega_a n_a + \omega_b n_b$ and eigenstates
$\vert n_a,n_b,n_c\rangle$ or equivalently $\vert N,J,J_z\rangle$
where the label $J=n_{ab}/2$ identifies the $SU(2)$ irrep.
These are states with well defined $n_a$, $n_b$ and $n_c=N-n_a-n_b$.
To create a pds
we add the term
\begin{equation} \label{h1}
H_1 = b^{\dagger}(b^\dagger a + b^\dagger c + a^\dagger b
+ c^\dagger b)b ~,
\end{equation}
which preserves the total boson number but not the individual boson
numbers, so it breaks the dynamical symmetry.
However states of the form $\vert n_a,n_b=0,n_c\rangle$ (or
equivalently $\vert N,J=n_a/2,J_z=-J\rangle$ ) with
$n_a=0,1,2,\ldots N$ are annihilated by $H_1$ and therefore remain
eigenstates of $H_0+BH_1$. The latter Hamiltonian
is not an $SU(2)$ scalar yet has a subset of $(N+1)$ ``special''
solvable states with $SU(2)$ symmetry, and therefore has pds.
There is one special state per $SU(2)$ irrep
$J=n_{a}/2$ (the lowest weight state in each case) with energy
$\omega_an_a$ independent of the parameter B.
Other eigenstates are mixed. Although $H_0$ and $H_1$ do not commute,
when acting on the ``special'' states they satisfy
\begin{equation} \label{comm}
\Bigl [H_0\, ,\, H_1\Bigr ]\vert n_a,n_b=0,n_c\rangle\; = 0 ~.
\end{equation}
To break the pds
we introduce a third interaction
\begin{equation} \label{h2}
H_2 = a^{\dagger}c + c^{\dagger}a + b^{\dagger}c + c^{\dagger}b ~.
\end{equation}
The complete Hamiltonian is then
\begin{equation} \label{htot}
H = H_0 + BH_1 + CH_2 ~.
\end{equation}
For $B=C=0$ we have the full dynamical symmetry; for
$B\neq 0,\,C=0$ we have partial dynamical symmetry
and for $C\neq 0$ we have neither.

The classical Hamiltonian ${\cal H}_{cl}$ is obtained from (\ref{htot})
by replacing $(a^\dagger,b^\dagger,c^\dagger)$ by complex c-numbers
$(\alpha^*,\beta^*,\gamma^*)$ and taking $N\rightarrow\infty$.
The latter
limit is obtained \cite{ibmchaos,mere}
by rescaling ${\bar B}=NB$, $\alpha\rightarrow \alpha/\sqrt{N}$ etc.
and considering the classical Hamiltonian per boson
${\cal H} = {\cal H}_{cl}/N$.
In the present model the latter has the form
\ba \label{cham}
{\cal H} &=&
{\cal H}_0 + \bar{B}{\cal H}_1 + C{\cal H}_2 ~.
\ea
Number conservation imposes a constraint
$\alpha^*\alpha+\beta^*\beta+\gamma^*\gamma=1$,
so that the phase space is compact and four-dimensional with a
volume $2\pi^2$. The total number of quantum states is $(N+1)(N+2)/2$.
Assigning, to leading order in $N$,
one state per $(2\pi\hbar)^2$ volume of phase space,
we identify
$\hbar=1/N$, so that the classical limit is
$N\rightarrow\infty$.

In all calculations reported below we take $A=0.8642$ and $N=60$.
As a first step, we fix ${\bar B}=0.5$ and vary $C$. For the classical
analysis we randomly sample the phase space and determine the fraction
$\mu$ of chaotic volume. For the quantum analysis we evaluated the energy
levels, calculated the nearest neighbors level spacing distribution of the
unfolded spectrum and fitted it to a Brody distribution \cite{brody}.
The Brody fit parameter $\omega$ is expected to be $0$ for integrable
systems (Poisson) and $1$ for chaotic systems (GOE). As shown in Fig.~1,
both of these measures indicate a suppression of chaos near $C=0$
similar to the results of Ref.~\cite{wal}.
To appreciate the strong effect of the pds
(at $C=0$)
on the underlying dynamics, it should be noted that the fraction of the
solvable states $\vert n_a,n_b=0,n_c\rangle$ is $2/(N+1)$, which approaches
zero in the classical limit.
To measure the extent to which each
eigenstate $|\Psi\rangle$ has $SU(2)$ symmetry,
we define variances $\sigma_{i}^2 =
\langle \Psi|\hat{n}_{i}^2| \Psi\rangle -
\langle \Psi|\hat{n}_{i}|\Psi\rangle^2$ $(i=a,b)$.
A state which belongs to just one irrep of $SU(2)$ (with well defined
$J,J_z$) has zero variances, while a mixed state
has large variances. These variances have the same physical content
as the entropies considered in Ref.\cite{wal}.
It is instructive to display the average $\langle\hat{n}_a\rangle$ and
variance of each state, as done in Fig.~2.
$SU(2)$ pds
is present in Fig.~2a ($B\neq0$, $C=0$),
Fig.~2b is a blow up of Fig~2a and in Fig.~2c
the symmetry is completely broken ($C\neq 0$).
In Figs.~2a-b we see states with zero variance. These
are just the special $N+1$ states ($n_b=0$) discussed before, which preserve
the $SU(2)$ symmetry. In addition, we see families of states with small
variance and small $\langle n_b\rangle$
which suggests that the presence of partial symmetry increases
the purity of states other than the special ones. By contrast, in
Fig.~2c we see no particular structure because of the destruction of
the pds
for $C\neq 0$.

Considerable insight is gained by examining the classical phase space
structure in terms of action-angle variables
$\alpha =\sqrt{J_a}\exp(-i\theta_a)$,
$\beta =\sqrt{J_b}\exp(-i\theta_b)$ and
$\gamma = \sqrt{J_c} = \sqrt{1-J_a-J_a}$.
The $\theta_a=-\pi/2$
Poincar\'e section is shown in Fig.~3 for energy $E=1.0$.
When $SU(2)$ pds
is present (${\bar B}\neq 0,\, C=0$)
we see in Figs.~3a-b a torus
with $J_b=0$, and additional perturbed tori in its neighborhood
(small $J_b$). This structure
is absent when the symmetry is completely broken ($C\neq 0$), as shown in
Fig.~3c. The features in Fig~3 persist also at other energies. To understand
them, we recall that for ${\bar B}=C=0$, the Hamiltonian (\ref{cham}) is integrable
and all trajectories wind around invariant tori.
By standard torus quantization (without turning points)
the actions are quantized as
$J_{i}=n_{i}\hbar=n_{i}/N$ $(i=a,b)$.
In the integrable limit quantum states are associated with toroidal
manifolds in phase space. In case of a partial symmetry 
(${\bar B}\neq 0,\,C=0$) we are led by analogy with Eq.~(\ref{comm}) 
to seek manifolds ${\cal M}$ in phase space on which
\begin{equation} \label{poiss}
\Bigl.\Bigl \{{\cal H}_0\, ,\,{\cal H}_1\Bigr \}\Bigr|_{\cal M}\; =0
\end{equation}
vanishes even though the Poisson bracket is not zero everywhere.
In addition, we demand
$\{\{{\cal H}_0\,,\,{\cal H}_1\},{\cal H}_0+{\cal H}_1\}|_{\cal M}=0$
(in analogy to the quantum relation
$[[\,H_0,H_1]\,,\,H_0+H_1]|n_a,n_b=0,n_c\rangle =0$)
so that a trajectory starting on
${\cal M}$ remains on ${\cal M}$. The solution to these conditions
is the manifold $J_b=\beta^*\beta=0$, which may be
interpreted as a (degenerate) torus of the ${\cal H}_0$
Hamiltonian. It is also a stable isolated periodic orbit of
${\cal H}_0+\bar{B}{\cal H}_1$.
Quantization of the torus with $J_b=0$ proceeds exactly
as before, so we correctly predict no change in the quantum energies
associated with it.
The manifold ${\cal M}$ ($J_b=0$) is the direct classical
analogue of the special quantum states $\vert n_b=0\rangle$.
It refers, however, to a region of phase space of measure zero, and so
cannot by itself explain the observed (global) suppression of chaos.
However, as suggested by Fig.~3, the presence of pds
induces a quasi-regular region foliated by tori in the vicinity
of the special torus. We can understand the dynamics on a finite measure of
phase space by performing a perturbative calculation in the neighbourhood
of ${\cal M}$.

For the classical perturbation calculation we set $C= 0$ in
Eq.~(\ref{cham}) and treat
$\bar{B}$ as an expansion parameter, assuming
$\bar{B}{\cal H}_1$ in Eq.~(\ref{cham}) is small in the neighbourhood
of the special periodic orbit. We seek a generating function for the
canonical transformation $({\bf J},\mbox{\boldmath $\theta$})\rightarrow
({\bf I},\mbox{\boldmath $\phi$})$ such that the
Hamiltonian depends only on the new actions ${\bf I}$ and not on the
new angles $\mbox{\boldmath $\phi$}$.
The first-order result in standard classical perturbation theory is
\begin{equation} \label{gener}
S({\bf I},\mbox{\boldmath $\theta$}) = {\bf I}\cdot\mbox{\boldmath
$\theta$} - 2\bar{B}I_b^{3/2}\left[\frac{I_a^{1/2}
\sin(\theta_a-\theta_b)}{(\Delta\omega)} -
\frac{I_c^{1/2}\sin\theta_b}{\omega_b}\right] ~,
\end{equation}
where $\Delta\omega=\omega_a - \omega_b$ and
${\bf J}=\frac{\partial S}{\partial {\bf I}}$,
$\mbox{\boldmath $\phi$} =
\frac{\partial S}{\partial \mbox{\boldmath $\theta$} }$. There is no first
order shift in the energy hence ${\cal H}(I) = \omega_a I_a +
\omega_b I_b$ and the calculation is valid for
$\bar{B}I_b^{3/2}\ll 1$.
The second order correction reproduces well
the perturbed tori on the Poincar\'e sections as shown in Fig.~3b.
The classical variances are defined by averaging with respect to the angles
($\theta_a,\theta_b$), e.g.
$\sigma_{a,cl}^2 = \langle J_a^{2}\rangle - \langle J_a\rangle^2$.
To the same order in perturbation theory,
$\langle J_{a}\rangle = I_a$, $\langle J_{b}\rangle = I_b$ and
the variances in $J_a$ and $J_b$ are
\begin{equation} \label{cdisp}
\sigma^2_{a,cl} = \frac{2\bar{B}^2I_b^3I_a}
{(\Delta\omega)^2}
\;\;\;\;\;\;\;\;\;\;\;\;
\sigma^2_{b,cl} =  2\bar{B}^2I_b^3
\left [\frac{I_a}{(\Delta\omega)^2}
+ \frac{I_c}{(\omega_b)^2}\right ] ~.
\end{equation}
The same calculation can also be done in
quantum perturbation theory with the result
\begin{eqnarray} \label{qdisp}
\sigma_a^2
& = & \frac{B^2n_b}{(\Delta\omega)^2}\left[(n_b-1)^2(n_a+1)
+n_b(n_b+1)n_a)\right] ~,
\nonumber\\
\sigma_b^2
& = & B^2n_b\left[\,\left(\frac{n_a}{(\Delta\omega)^2}
+ \frac{n_c}{(\omega_b)^2}\right)(2n_b^2-n_b+1)
+ \left(\frac{1}{(\Delta\omega)^2}+\frac{1}{(\omega_b)^2}\right)
(n_b-1)^2\,\right] ~.
\end{eqnarray}
In the semiclassical limit, $n_a,n_b,n_c\gg 1$,
Eqs.~(\ref{cdisp}) and (\ref{qdisp}) agree to leading order.
In Fig.~2b we show the results (denoted by ($+$) ) of the quantum
perturbation theory (to order $B^5$).
We see that the first few families of states are reproduced.
It is these states which we can recover from perturbation
theory and whose approximate symmetry is induced by the
symmetry of the special states.

The following physical picture emerges from the foregoing analysis.
Near the special orbit, there are KAM tori,
some of which are quantized. The quantum eigenstates lie on these
tori, so knowing the classical variance of the actions of the tori tells us
the variances of the states themselves, in the semiclassical limit.
Large variances indicate the extent to which the corresponding states fail
to respect the symmetry. This provides a measure for a separation of regular
and irregular levels, as conceived by Percival \cite{perc} and which
recently gained numerical support \cite{btu90,lirob}.
In the present model the
quantum states can be grouped into three classes:
i) the special states, which observe the symmetry;
ii) the ``almost special states'' which are accessible by perturbation
theory; iii) the rest of the states, which are mixed.
As in \cite{btu90}, the frontier between regular states (sets (i) and (ii) )
and irregular states (set (iii) ) is not sharp.

In summary, we have considered the effect of partial dynamical symmetry
(pds) on the quantum and classical dynamics of a mixed system. At the
quantum level, pds
by definition implies the existence
of a ``special'' subset of  states, which observe the symmetry. The pds
affects the purity
of other states in the system; in particular, neighboring states,
accessible by perturbation theory,
possess approximately good symmetry. Analogously,
at the classical level, the region of phase space near the ``special''
torus also has toroidal structure. As a consequence
of having pds,
a finite region of phase space is
regular and a finite fraction of states is approximately ``special''.
This clarifies the observed suppression of chaos. Based on these
arguments and the numerical results of Ref. \cite{wal}, we anticipate the
suppression of chaos to persist in higher dimensional systems with partial
dynamical symmetries.

This research was partly supported by the EU Human Capital and Mobility
Programme, by a Canadian NSERC fellowship (N.D.W.) and by
the Israel Science Foundation administered by the Israel Academy of Sciences
and Humanities (A.L.). N.D.W. thanks Y. Alhassid and S. Creagh for
useful discussions.

\begin{figure}
\caption{Classical ($\mu$) and quantum ($\omega$) measures of chaos
(denoted by ($\bullet$) and ($\times$) respectively) versus $C$
for the Hamiltonian (6) with ${\bar B}=0.5$.}
\end{figure}

\begin{figure}
\caption{The values of $\langle n_a\rangle$ and of the variance $\sigma_b$
(denoted by $\diamond$) of each eigenstate of the Hamiltonian (6).
(a)~${\bar B}=0.5$, $C=0$ (partial dynamical symmetry).
(b)~a blow up of (a) with superimposed results (denoted by ($+$) )
of quantum perturbation theory.
The families of states with low $\sigma_b$
have small values of $\langle n_b\rangle$.
(c)~${\bar B}=0.3$, $C=0.5$ (broken symmetry).}
\end{figure}

\begin{figure}
\caption{
Poincar\'{e} sections $J_b$ versus $\theta_b$ at energy $E=1.0$.
(a)~${\bar B}=0.5$, $C=0$ (partial dynamical symmetry).
(b)~a blow up of (a) with superimposed results (dashed curves)
of classical perturbation theory.
(c)~${\bar B}=0.3$, $C=0.5$ (broken symmetry).}
\end{figure}
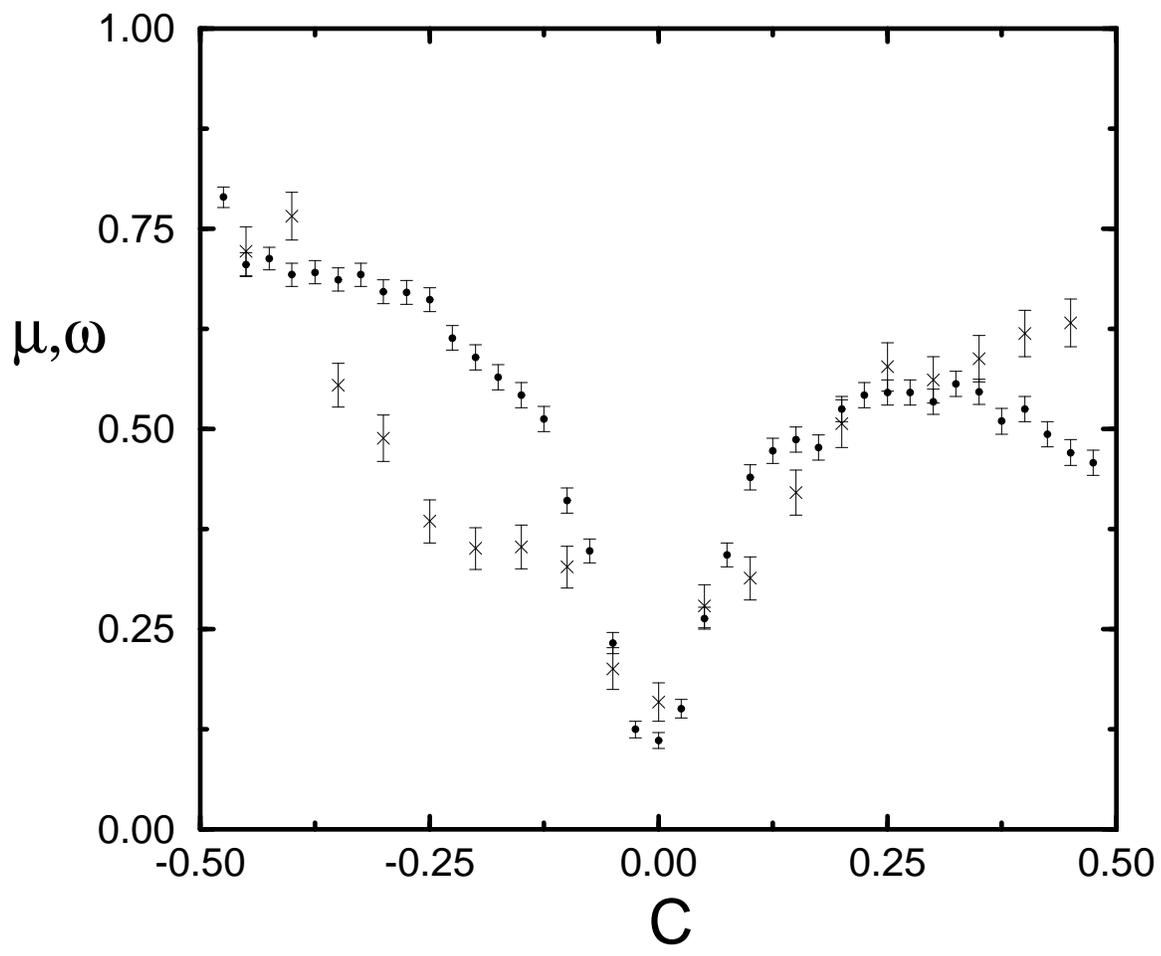
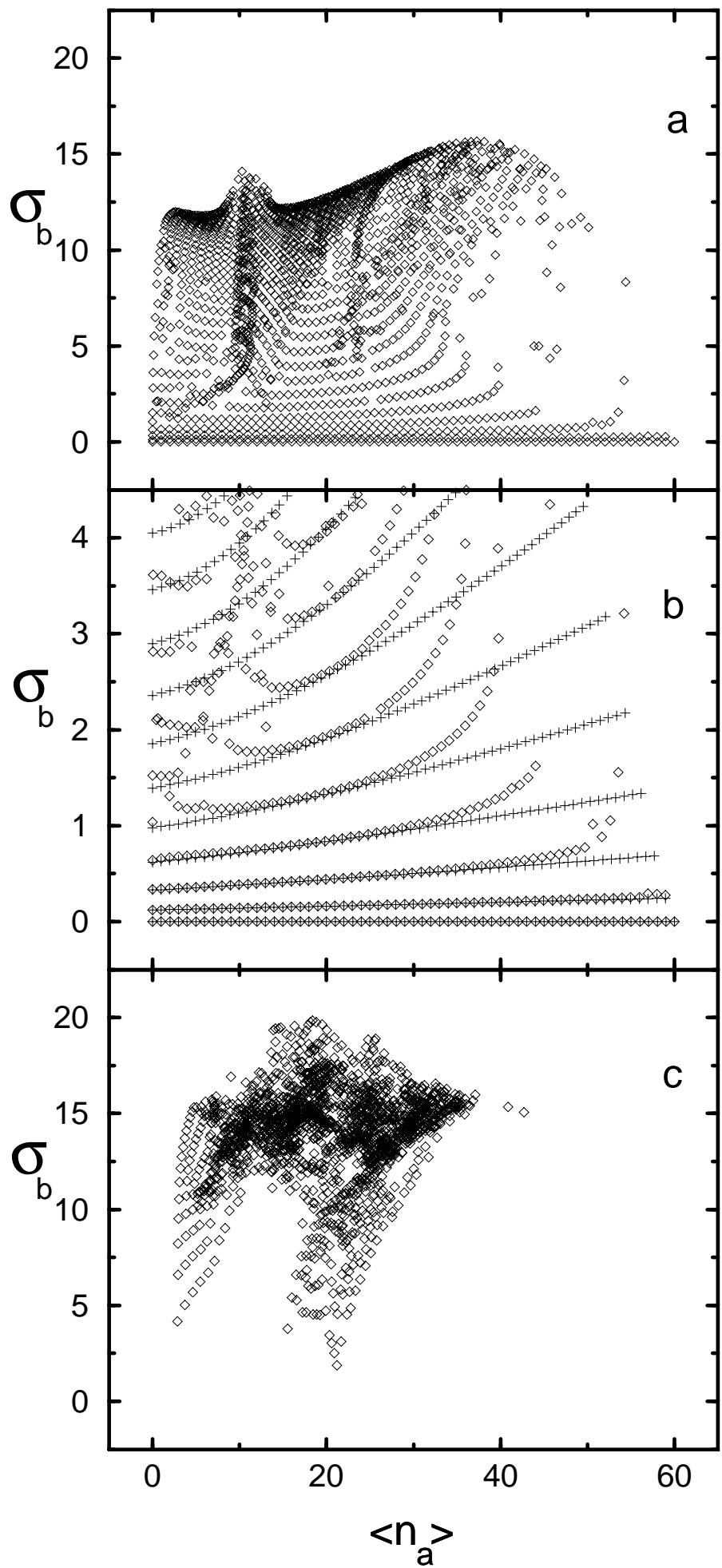
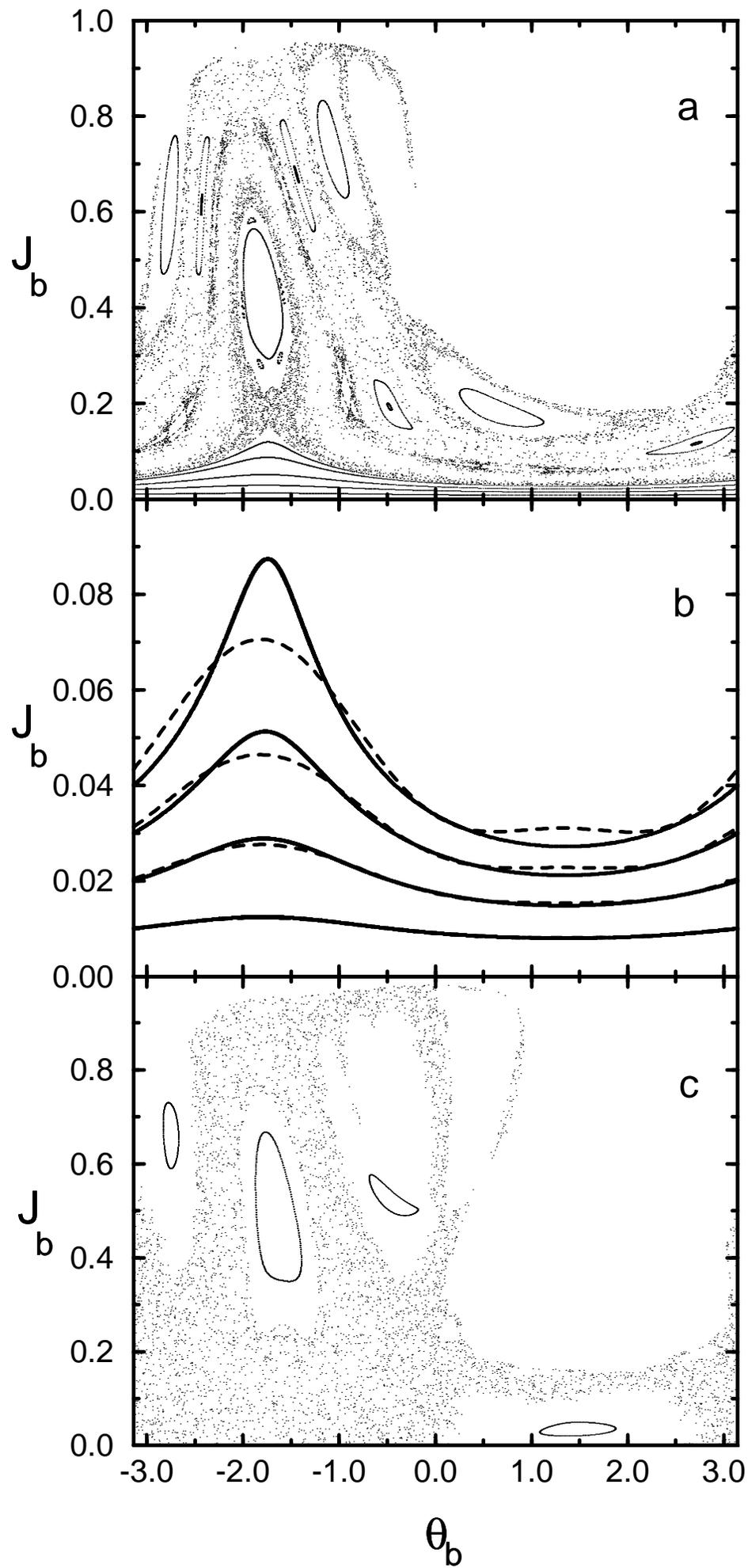
\end{document}